\documentclass[10pt,a4paper]{article}
\usepackage{amsmath}
\usepackage{amssymb}
\usepackage{amsfonts}
\usepackage{amstext}
\usepackage{amsbsy}
\usepackage[mathscr]{eucal}
\usepackage{graphicx}
\usepackage{color}
\newcommand{\beq}{\begin{equation}}
\newcommand{\eeq}{\end{equation}}
\newcommand{\beqa}{\begin{eqnarray}}
\newcommand{\eeqa}{\end{eqnarray}}
\newcommand{\ket}[1]{| #1 \rangle}
\newcommand{\bra}[1]{\langle #1 |}

\title{\Large\textbf{Concurrence for multipartite states}}

\author{\textit{Hoshang Heydari}\\{\small\emph{Institute of Quantum
Science, Nihon University,}}\\ \small{\emph{1-8 Kanda-Surugadai,
Chiyoda-ku, Tokyo 101-8308, Japan}}}
\begin{document}

\maketitle

\begin{abstract}
We  construct a generalized concurrence for general multipartite
states based on local W-class and  GHZ-class operators. We
explicitly construct the corresponding concurrence for three-partite
states. The construction of the concurrence is interesting since it
is based on local operators.
\end{abstract}

\section{Introduction}
Concurrence is one of the most applied measure of entanglement. In recent years there have been some proposals to
generalize this measure of entanglement to general multipartite
states \cite{Albeverio,Mintert}.
 Recently, we have also defined
concurrence classes for multi-qubit mixed states  based on an
orthogonal complement of a positive operator valued measure (POVM)
on quantum phase \cite{Hosh5}. Moreover, we have constructed
different concurrence classes for general pure multipartite states
in \cite{Hosh6}. In this paper, we will construct generalized
concurrence for pure general multipartite states based on the
complement of a POVM on quantum phase. However, this measure is not
equal to our concurrence classes, where we have added these
concurrence classes and then took the square root of them. But by
rewriting our linear operators as sums and take the expectation
value of each of these operators, we are able to construct a general
formula for concurrence.
 We will consider  a general, multipartite quantum
system with $m$ subsystems
$\mathcal{Q}=\mathcal{Q}_{m}(N_{1},N_{2},\ldots,N_{m})$, denoting
its general state as $
\ket{\Phi}=\sum^{N_{1}}_{l_{1}=1}\cdots\sum^{N_{m}}_{l_{m}=1}
\alpha_{l_{1},l_{2},\ldots,l_{m}} \ket{l_{1},l_{2},\ldots,l_{m}} $.
Moreover, let
$\rho_{\mathcal{Q}}=\sum^{\mathrm{N}}_{n=1}p_{n}\ket{\Phi_{n}}\bra{\Phi_{n}}$,
for all $0\leq p_{n}\leq 1$ and $\sum^{\mathrm{N}}_{n=1}p_{n}=1$,
denote a density operator acting on the Hilbert space $
\mathcal{H}_{\mathcal{Q}}=\mathcal{H}_{\mathcal{Q}_{1}}\otimes
\mathcal{H}_{\mathcal{Q}_{2}}\otimes\cdots\otimes\mathcal{H}_{\mathcal{Q}_{m}},
$ where the dimension of the $j$th Hilbert space is given  by
$N_{j}=\dim(\mathcal{H}_{\mathcal{Q}_{j}})$. Finally, let us
introduce a complex conjugation operator $\mathcal{C}_{m}$ that acts
on a general multipartite state $\ket{\Phi}$  as $
\mathcal{C}_{m}\ket{\Phi}=\sum^{N_{1}}_{l_{1}=1}\cdots\sum^{N_{m}}_{l_{m}=1}
\alpha^{*}_{l_{1},l_{2},\ldots,l_{m}}
\ket{l_{1},l_{2},\ldots,l_{m}}.$


\section{General multipartite
states}\label{conclass} In this section, we will construct
concurrence for general pure multipartite states
$\mathcal{Q}^{p}_{m}(N_{1},\ldots,N_{m})$. In our construction, we
will use linear operators that are constructed by the orthogonal
complement of POVM on quantum phase \cite{Hosh5,Hosh6}. In order to
simplify our presentation, we will use $\Lambda_{m}=k_{1},l_{1};$
$\ldots;k_{m},l_{m}$ as an abstract multi-index notation.  In the
$m$-partite case, the off-diagonal elements of the matrix
corresponding to
\begin{eqnarray}\nonumber
\widetilde{\Delta}_\mathcal{Q}(\varphi_{\mathcal{Q}_{1};k_{1},l_{1}},\ldots,
\varphi_{\mathcal{Q}_{m};k_{m},l_{m}})&=&
\widetilde{\Delta}_{\mathcal{Q}_{1}}(\varphi_{\mathcal{Q}_{1};k_{1},l_{1}})
\otimes\cdots
\otimes\widetilde{\Delta}_{\mathcal{Q}_{m}}(\varphi_{\mathcal{Q}_{m};k_{m},l_{m}}),
\\
\end{eqnarray}
where the orthogonal complement of our POVM
\begin{eqnarray}
&&\Delta(\varphi_{\mathcal{Q}_{j};k_{j},l_{j}})=
\sum^{N_{j}}_{l_{j},k_{j}=1}
e^{i\varphi_{k_{j},l_{j}}}\ket{k_{j}}\bra{l_{j}}
\end{eqnarray} is given by
$\widetilde{\Delta}_{\mathcal{Q}_{j}}(\varphi_{\mathcal{Q}_{j};k_{j},l_{j}})=\mathcal{I}_{N_{j}}-
\Delta_{\mathcal{Q}_{j}}(\varphi_{\mathcal{Q}_{j};k_{j},l_{j}})$.
$\mathcal{I}_{N_{j}}$ is the $N_{j}$-by-$N_{j}$ identity matrix for
subsystem $j$.
 $\widetilde{\Delta}_\mathcal{Q}(\varphi_{\mathcal{Q}_{1};k_{1},l_{1}},\ldots,
\varphi_{\mathcal{Q}_{m};k_{m},l_{m}})$ has phases that are sums or
differences of phases originating from two and $m$ subsystems. That
is, in the latter case the phases of
$\widetilde{\Delta}_\mathcal{Q}(\varphi_{\mathcal{Q}_{1};k_{1},l_{1}},\ldots,
\varphi_{\mathcal{Q}_{m};k_{m},l_{m}})$ take the form
$(\varphi_{\mathcal{Q}_{1};k_{1},l_{1}}\pm\varphi_{\mathcal{Q}_{2};k_{2},l_{2}}
\pm\ldots\pm\varphi_{\mathcal{Q}_{m};k_{m},l_{m}})$ and
identification of these joint phases makes our distinguishing
possible. Thus, we can define linear operators for the
$\mathrm{W}^{m}$ class which are sums and differences of phases of
two subsystems, i.e.,
$(\varphi_{\mathcal{Q}_{r_{1}};k_{r_{1}},l_{r_{1}}}
\pm\varphi_{\mathcal{Q}_{r_{2}};k_{r_{2}},l_{r_{2}}})$. That is, for
the $\mathrm{W}^{m}$ class we have
\begin{eqnarray}
 \widetilde{\Delta}^{
\mathrm{W}^{m}_{\Lambda_{m}}}_{\mathcal{Q}_{r_{1},r_{2}}(N_{r_{1}},N_{r_{2}})}
&=&\mathcal{I}_{N_{1}} \otimes\cdots
\otimes\widetilde{\Delta}_{\mathcal{Q}_{r_{1}}}
(\varphi^{\frac{\pi}{2}}_{\mathcal{Q}_{r_{1}};k_{r_{1}},l_{r_{1}}})\\\nonumber&&
\otimes\cdots\otimes \widetilde{\Delta}_{\mathcal{Q}_{r_{2}}}
(\varphi^{\frac{\pi}{2}}_{\mathcal{Q}_{r_{2}};k_{r_{2}},l_{r_{2}}})\otimes\cdots\otimes
\mathcal{I}_{N_{m}}.
\end{eqnarray}
Next, we could write the linear operator
$\widetilde{\Delta}^{\mathrm{W}^{m}_{\Lambda_{m}}}_{\mathcal{Q}_{r_{1},r_{2}}(N_{r_{1}},N_{r_{2}})}$
as a direct sum of the upper and lower anti-diagonal
\begin{eqnarray}
 \widetilde{\Delta}^{
\mathrm{W}^{m}_{\Lambda_{m}}}_{\mathcal{Q}_{r_{1},r_{2}}(N_{r_{1}},N_{r_{2}})}
&=&\mathfrak{U}\widetilde{\Delta}^{
\mathrm{W}^{m}_{\Lambda_{m}}}_{\mathcal{Q}_{r_{1},r_{2}}(N_{r_{1}},N_{r_{2}})}+\mathfrak{L}\widetilde{\Delta}^{
\mathrm{W}^{m}_{\Lambda_{m}}}_{\mathcal{Q}_{r_{1},r_{2}}(N_{r_{1}},N_{r_{2}})}.
\end{eqnarray}
The  set of  linear operators for the $\mathrm{W}^{m}$ classes gives
the $\mathrm{W}^{m}$ class concurrence.

 For the $\mathrm{GHZ}^{m}$ class, we define linear
operators based on our POVM which are sums and differences of phases
of $m$-subsystems, i.e.,
$(\varphi_{\mathcal{Q}_{r_{1}};k_{r_{1}},l_{r_{1}}}
\pm\varphi_{\mathcal{Q}_{r_{2}};k_{r_{2}},l_{r_{2}}}\pm
\ldots\pm\varphi_{\mathcal{Q}_{m};k_{m},l_{m}})$. That is, for the
$\mathrm{GHZ}^{m}$ class we have
\begin{eqnarray}
 \widetilde{\Delta}^{
\mathrm{GHZ}^{m}_{\Lambda_{m}}}_{\mathcal{Q}_{r_{1},r_{2}}(N_{r_{1}},N_{r_{2}})}
&=&\widetilde{\Delta}_{\mathcal{Q}_{1}}
(\varphi^{\pi}_{\mathcal{Q}_{1};k_{1},l_{1}})\otimes\cdots
\otimes\widetilde{\Delta}_{\mathcal{Q}_{r_{1}}}
(\varphi^{\frac{\pi}{2}}_{\mathcal{Q}_{r_{1}};k_{r_{1}},l_{r_{1}}})\\\nonumber&&
\otimes\cdots\otimes \widetilde{\Delta}_{\mathcal{Q}_{r_{2}}}
(\varphi^{\frac{\pi}{2}}_{\mathcal{Q}_{r_{2}};k_{r_{2}},l_{r_{2}}})\otimes\cdots\otimes
\widetilde{\Delta}_{\mathcal{Q}_{m}}
(\varphi^{\pi}_{\mathcal{Q}_{m};k_{m},l_{m}}).
\end{eqnarray}
where by choosing $\varphi^{\pi}_{\mathcal{Q}_{j};k_{j},l_{j}}=\pi$
for all $k_{j}<l_{j}, ~j=1,2,\ldots,m$, we get an operator which has
the structure of the Pauli operator $\sigma_{x}$ embedded in a
higher-dimensional Hilbert space and coincides with $\sigma_{x}$ for
a single-qubit. There are $\frac{m(m-1)}{2} $ linear operators for
the $\mathrm{GHZ}^{m}$ class.

Next, we write the linear operators for the $\mathrm{GHZ}^{m}$ class
as
\begin{eqnarray}
 \widetilde{\Delta}^{
\mathrm{GHZ}^{m}_{\Lambda_{m}}}_{\mathcal{Q}_{r_{1},r_{2}}(N_{r_{1}},N_{r_{2}})}
&=&\mathfrak{P}_{1}\widetilde{\Delta}^{
\mathrm{GHZ}^{m}_{\Lambda_{m}}}_{\mathcal{Q}_{r_{1},r_{2}}(N_{r_{1}},N_{r_{2}})}+\mathfrak{P}_{2}\widetilde{\Delta}^{
\mathrm{GHZ}^{m}_{\Lambda_{m}}}_{\mathcal{Q}_{r_{1},r_{2}}(N_{r_{1}},N_{r_{2}})}+\ldots,
\end{eqnarray}
 where the
operators $\mathfrak{P}_{i}\widetilde{\Delta}^{
\mathrm{GHZ}^{m}_{\Lambda_{m}}}_{\mathcal{Q}_{r_{1},r_{2}}(N_{r_{1}},N_{r_{2}})}$
are constructed by pairing of elements of the POVM with sums and
differences of quantum phases. For higher dimensional quantum
systems, it is difficult to write
$\widetilde{\Delta}^{\mathrm{GHZ}^{m}_{\Lambda_{m}}}_{\mathcal{Q}_{r_{1},r_{2}}(N_{r_{1}},N_{r_{2}})}$
in terms of $\mathfrak{P}_{i}\widetilde{\Delta}^{
\mathrm{GHZ}^{m}_{\Lambda_{m}}}_{\mathcal{Q}_{r_{1},r_{2}}(N_{r_{1}},N_{r_{2}})}$.
However, we will give an explicit expression for general
three-partite states in the next section.
 Moreover, we define the
linear operators for the $\mathrm{GHZ}^{m-1}$ class of $m$-partite
states based on our POVM which are sums and differences of phases of
$m-1$-subsystems, i.e.,
$(\varphi_{\mathcal{Q}_{r_{1}};k_{r_{1}},l_{r_{1}}}
\pm\varphi_{\mathcal{Q}_{r_{2}};k_{r_{2}},l_{r_{2}}}
\pm\ldots\varphi_{\mathcal{Q}_{m-1};k_{m-1},l_{m-1}}\pm\varphi_{\mathcal{Q}_{m-1};k_{m-1},l_{m-1}})$.
That is, for the $\mathrm{GHZ}^{m-1}$ class we have

\begin{eqnarray}
\widetilde{\Delta}^{
\mathrm{GHZ}^{m-1}_{\Lambda_{m}}}_{\mathcal{Q}_{r_{1}r_{2},r_{3}}(N_{r_{1}},N_{r_{2}})}
&=& \widetilde{\Delta}_{\mathcal{Q}_{r_{1}}}
(\varphi^{\frac{\pi}{2}}_{\mathcal{Q}_{r_{1}};k_{r_{1}},l_{r_{1}}})
\otimes\widetilde{\Delta}_{\mathcal{Q}_{r_{2}}}
(\varphi^{\frac{\pi}{2}}_{\mathcal{Q}_{r_{2}};k_{r_{2}},l_{r_{2}}})
\otimes\\\nonumber&&\widetilde{\Delta}_{\mathcal{Q}_{r_{3}}}
(\varphi^{\pi}_{\mathcal{Q}_{r_{3}};k_{r_{3}},l_{r_{3}}})
\otimes\cdots
\otimes\\\nonumber&&\widetilde{\Delta}_{\mathcal{Q}_{m-1}}
(\varphi^{\pi}_{\mathcal{Q}_{m-1};k_{r_{m-1}},l_{r_{m-1}}})\otimes\mathcal{I}_{N_{m}}
,
\end{eqnarray}
where $1\leq r_{1}<r_{2}<\cdots<r_{m-1}<m$. Note that we need to
write these operators also as direct sums as we did for
$\mathrm{GHZ}^{m}$ class since they belong to the same operator
class. Then, for a general pure state
 let
\begin{eqnarray}
\mathcal{C}(\mathcal{Q}^{W^{m}}_{r_{1},r_{2}}(N_{r_{1}},N_{r_{2}}))&=&
    \sum_{\forall k_{j},l_{j}}(
    \left|\langle \Phi\ket{\mathfrak{U}\widetilde{\Delta}^{
W^{m}_{\Lambda_{m}}}_{\mathcal{Q}_{r_{1},r_{2}}(N_{r_{1}},N_{r_{2}})}\mathcal{C}_{m}\Phi}
\right|^{^{2}}\\\nonumber&&+ \left|\langle
\Phi\ket{\mathfrak{L}\widetilde{\Delta}^{
W^{m}_{\Lambda_{m}}}_{\mathcal{Q}_{r_{1},r_{2}}(N_{r_{1}},N_{r_{2}})}\mathcal{C}_{m}\Phi}
\right|^{^{2}}),
\end{eqnarray}
\begin{eqnarray}
   \mathcal{C}(\mathcal{Q}^{GHZ^{m}}_{r_{1},r_{2}}(N_{r_{1}},N_{r_{2}}))&=&
    \sum_{\forall k_{j},l_{j}}\sum_{i\geq m-2}
    \left|\langle \Phi\ket{\mathfrak{P}_{i}\widetilde{\Delta}^{
GHZ^{m}_{\Lambda_{m}}}_{\mathcal{Q}_{r_{1},r_{2}}(N_{r_{1}},N_{r_{2}})}\mathcal{C}_{m}\Phi}
\right|^{^{2}}
\end{eqnarray}
and e.g.,
\begin{eqnarray}
   \mathcal{C}(\mathcal{Q}^{GHZ^{m-1}}_{r_{1}r_{2},r_{3}}(N_{r_{1}},N_{r_{2}}))&=&
    \sum_{\forall k_{j},l_{j}}\sum_{i\geq m-3}
    \left|\langle \Phi\ket{\mathfrak{P}_{i}\widetilde{\Delta}^{
GHZ^{m-1}_{\Lambda_{m}}}_{\mathcal{Q}_{r_{1}r_{2},r_{3}}(N_{r_{1}},N_{r_{2}})}\mathcal{C}_{m}\Phi}
\right|^{^{2}}.
\end{eqnarray}
Then the concurrence is defined by adding these terms and the take
square root of them as follows
\begin{eqnarray}
\mathcal{C}(\mathcal{Q}^{p}_{m}(N_{1},\ldots,N_{m}))&=&
    (\mathcal{N}_{m}\{\sum^{m}_{r_{2}>r_{1}=1}\mathcal{C}(\mathcal{Q}^{W^{m}}_{r_{1},r_{2}}
    (N_{r_{1}},N_{r_{2}}))\\\nonumber&&
    +\sum^{m}_{r_{2}>r_{1}=1}\mathcal{C}(\mathcal{Q}^{GHZ^{m}}_{r_{1},r_{2}}
    (N_{r_{1}},N_{r_{2}}))\\\nonumber&&
    +\sum^{m}_{r_{3}>r_{2}>r_{1}=1}\mathcal{C}(\mathcal{Q}^{GHZ^{m-1}}_{r_{1}r_{2},r_{3}}
    (N_{r_{1}},N_{r_{2}}))+\ldots\})^{1/2},
\end{eqnarray}
where $\mathcal{N}_{m}$ is a normalization constant. Note that for
three-partite states our concurrence consists of two parts
$\mathcal{C}(\mathcal{Q}^{W^{3}}_{r_{1},r_{2}}(N_{r_{1}},N_{r_{2}}))$
and
$\mathcal{C}(\mathcal{Q}^{GHZ^{3}}_{r_{1},r_{2}}(N_{r_{1}},N_{r_{2}}))$
which we will discuss in the next section. However, for four-partite
states we have
$\mathcal{C}(\mathcal{Q}^{W^{3}}_{r_{1},r_{2}}(N_{r_{1}},N_{r_{2}}))$,
$\mathcal{C}(\mathcal{Q}^{GHZ^{3}}_{r_{1},r_{2}}(N_{r_{1}},N_{r_{2}}))$,
and
$\mathcal{C}(\mathcal{Q}^{GHZ^{3}}_{r_{1}r_{2},r_{3}}(N_{r_{1}},N_{r_{2}}))$.
Moreover, we can in principe define a concurrence for arbitrary
multipartite states as
\begin{eqnarray}
\mathcal{C}(\mathcal{Q}_{m}(N_{1},\ldots,N_{m}))&=&
\inf_{\Phi}\mathcal{C}(\mathcal{Q}^{p}_{m}(N_{1},\ldots,N_{m})).
\end{eqnarray}
However, to evaluate it one needs to find a pure decomposition of
density matrix of a given multipartite state which is a very
difficult task.

\section{General pure three-partite states}
\label{threepart} In this section, as an illustrative example,  we
will construct concurrence for pure three-partite quantum system
$\mathcal{Q}^{p}_{3}(N_{1},N_{2},N_{3})$ based on the orthogonal
complement of our POVM.
 For three-partite states, we have two
different joint phases in our POVM, those which are sums and
differences of phases of two subsystems, i.e.,
$(\varphi_{\mathcal{Q}_{1};k_{1},l_{1}}\pm\varphi_{\mathcal{Q}_{2};k_{2},l_{2}})$
and those which are sums and differences of phases of three
subsystems, i.e.,
$(\varphi_{\mathcal{Q}_{1};k_{1},l_{1}}\pm\varphi_{\mathcal{Q}_{2};k_{2},l_{2}}
\pm\varphi_{\mathcal{Q}_{3};k_{3},l_{3}})$. The first one identifies
the $\mathrm{W}^{3}$ class operator and the second one identifies
the $\mathrm{GHZ}^{3}$ class operator.
For  the $\mathrm{W}^{3}$ class, we have
\begin{eqnarray}\nonumber
    &&\mathcal{C}(\mathcal{Q}^{\mathrm{W}^{3}}_{3}(N_{1},N_{2},N_{3}))=\\\nonumber&&\sum^{N_{1}}_{l_{1}>k_{1}=1}
\sum^{N_{2}}_{l_{2}>k_{2}=1} \sum^{N_{3}}_{k_{3}=l_{3}=1}|
\alpha_{k_{1},l_{2},k_{3}}\alpha_{l_{1},k_{2},l_{3}}-\alpha_{k_{1},k_{2},k_{3}}\alpha_{l_{1},l_{2},l_{3}}
|^{2}
\\\nonumber&&+\sum^{N_{1}}_{l_{1}>k_{1}=1}
\sum^{N_{3}}_{l_{3}>k_{3}=1}
\sum^{N_{2}}_{k_{2}=l_{2}=1}|\alpha_{k_{1},k_{2},l_{3}}\alpha_{l_{1},l_{2},k_{3}}
-\alpha_{k_{1},k_{2},k_{3}}\alpha_{l_{1},l_{2},l_{3}} |^{2} \\&&+
\sum^{N_{2}}_{l_{2}>k_{2}=1} \sum^{N_{3}}_{l_{3}>k_{3}=1}
\sum^{N_{1}}_{k_{1}=l_{1}=1}|
\alpha_{k_{1},k_{2},l_{3}}\alpha_{l_{1},l_{2},k_{3}}-\alpha_{k_{1},k_{2},k_{3}}\alpha_{l_{1},l_{2},l_{3}}
|^{2},
\end{eqnarray}
and for the $\mathrm{GHZ}^{3}$ class,  we have
\begin{eqnarray}\label{cone}
  && \mathcal{C}(\mathcal{Q}^{\mathrm{GHZ}^{3}}_{3}(N_{1},N_{2},N_{3}))=
    \sum^{N_{1}}_{l_{1}>k_{1}=1}
\sum^{N_{2}}_{l_{2}>k_{2}=1}
\sum^{N_{3}}_{l_{3}>k_{3}=1}[\\\nonumber&&
|\alpha_{k_{1},l_{2},l_{3}}\alpha_{l_{1},k_{2},k_{3}}
-\alpha_{k_{1},k_{2},k_{3}}\alpha_{l_{1},l_{2},l_{3}}|^{2}
+|\alpha_{k_{1},l_{2},k_{3}}\alpha_{l_{1},k_{2},l_{3}}
-\alpha_{k_{1},k_{2},l_{3}}\alpha_{l_{1},l_{2},k_{3}}|^{2}\\\nonumber&&
+
|\alpha_{k_{1},k_{2},l_{3}}\alpha_{l_{1},l_{2},k_{3}}-\alpha_{k_{1},l_{2},l_{3}}\alpha_{l_{1},k_{2},k_{3}}|^{2}+
|\alpha_{k_{1},l_{2},k_{3}}\alpha_{l_{1},k_{2},l_{3}}
-\alpha_{k_{1},k_{2},k_{3}}\alpha_{l_{1},l_{2},l_{3}}|^{2}\\\nonumber&&+
|
\alpha_{k_{1},l_{2},k_{3}}\alpha_{l_{1},k_{2},l_{3}}-\alpha_{k_{1},l_{2},l_{3}}\alpha_{l_{1},k_{2},k_{3}}|^{2}
+|\alpha_{k_{1},k_{2},l_{3}}\alpha_{l_{1},l_{2},k_{3}}
-\alpha_{k_{1},k_{2},k_{3}}\alpha_{l_{1},l_{2},l_{3}}|^{2}].
\end{eqnarray}
Note that these expressions are not equal to our $\mathrm{W}$ class and $\mathrm{GHZ}$ class concurrences
constructed in \cite{Hosh6}, where we have constructed our
concurrences classes based on direct use of two class of operators.  Thus
the concurrence for a general pure three-partite state is give by
\begin{eqnarray}
    \mathcal{C}(\mathcal{Q}^{p}_{3}(N_{1},N_{2},N_{3}))&=&(\mathcal{N}_{3}
    [\mathcal{C}(\mathcal{Q}^{\mathrm{W}^{3}}_{3}(N_{1},N_{2},N_{3}))\\\nonumber&&+
    \mathcal{C}(\mathcal{Q}^{\mathrm{GHZ}^{3}}_{3}(N_{1},N_{2},N_{3}))])^{1/2}.
\end{eqnarray}
This concurrence also coincides with the generalized concurrence for
three-partite states\cite{Albeverio}. Moreover, for $m$-partite
states with $m\geq3$, our concurrence is not the equal to
concurrence tensor \cite{Hosh4}.
\begin{flushleft}
\emph{Acknowledgments:} The author acknowledges useful comments from
Jonas S\"{o}derholm. The  author gratefully acknowledges the
financial support of the Japan Society for the Promotion of Science
(JSPS).
\end{flushleft}


\end{document}